\documentclass[preprint,showpacs,preprintnumbers,amsmath,amssymb]{revtex4}
\usepackage{graphics,epsfig,subfigure}
\usepackage{color}

\begin{document}
\renewcommand{\baselinestretch}{1.3}
\newcommand\beq{\begin{equation}}
\newcommand\eeq{\end{equation}}
\newcommand\beqn{\begin{eqnarray}}
\newcommand\eeqn{\end{eqnarray}}
\newcommand\nn{\nonumber}
\newcommand\fc{\frac}
\newcommand\lt{\left}
\newcommand\rt{\right}
\newcommand\pt{\partial}

\title{Scalar perturbations of Eddington-inspired Born-Infeld braneworld}
\author{Ke Yang$^a$\footnote{keyang@swu.edu.cn},
        Yu-Xiao Liu$^b$\footnote{liuyx@lzu.edu.cn, corresponding author},
        Bin Guo$^c$\footnote{binguo@phas.ubc.ca}
        and Xiao-Long Du$^d$\footnote{xiaolong@astro.physik.uni-goettingen.de}
}

\affiliation{
$^{a}$School of Physical Science and Technology, Southwest University, Chongqing 400715, China\\
$^{b}$Institute of Theoretical Physics, Lanzhou University, Lanzhou 730000, China\\
$^{c}$Department of Physics and Astronomy, University of British Columbia, Vancouver, British Columbia V6T 1Z1, Canada \\
$^{d}$Institut f\"{u}r Astrophysik, Universit\"at G\"ottingen,
Friedrich-Hund-Platz 1, D-37075 G\"{o}ttingen, Germany
}

\begin{abstract}
We consider the scalar perturbations of Eddington-inspired Born-Infeld braneworld models in this paper. The dynamical equation for the physical propagating degree of freedom $\xi(x^\mu,y)$ is achieved by using the Arnowitt-Deser-Misner decomposition method: $F_1(y) {\partial_y^2 \xi} + F_2(y){\partial_y \xi  } + {\partial^{\mu}\partial_{\mu}}\xi=0$. We conclude that the solution is tachyonic-free and stable under scalar perturbations for $F_1(y)>0$ but unstable for $F_1(y)< 0$. The stability of a known analytic domain wall solution with the warp factor given by $a(y)= \text{sech}^{\frac{3}{{4p}}}(ky)$ is analyzed and it is shown that only the solution for $0<p<\sqrt{8/3}$ is stable.
\end{abstract}

% \Keywords{ }


%\pacs{04.50.Kd, 98.80.-k}


%11.10.Kk Field theories in dimensions other than four (see also 04.50.-h Higher-dimensional gravity and other theories of gravity; 04.60.Kz Lower dimensional models; minisuperspace models in general relativity and gravitation)

%04.50.Kd 	Modified theories of gravity

%04.50.-h Higher-dimensional gravity and other theories of gravity
%         (see also 11.25.Mj Compactification and four-dimensional models, 11.25.Uv D branes)

% 04.50.+h Gravity in more than four dimensions, Kaluza-Klein theory,
           % unified field theories, alternative theories of gravity
           %(see also 11.25.M Compactification and four-dimensional models), dilaton gravity
% 11.27.+d Extended classical solutions; cosmic strings,
           %domain walls, texture (see 98.80.C in cosmology)
%98.80.-k Cosmology
%\keywords{}

\maketitle


%%%%%%%%%%%%%%%%%%%%%%%%%%%%%%%%%%%%%%%%%%%%%%%%%%%%%%%%%%%%%%%%%%%%%%%%%%%%%%%%%%%%%%%

\section{Introduction}

It is widely accepted that general relativity should be modified in the ultraviolet regime, since the theory suffers from various troublesome problems, such as the inevitable singularities in cosmology and gravitational collapse \cite{Hawking1970}, and quantizing the general relativity leads to a nonrenormalizable quantum theory due to a dimensionful Newton's constant \cite{Hooft1974}.  So some modified gravities may unveil the corner of the unknown quantum gravity theory. It is well known that Born-Infeld electrodynamics proposed in 1934 can remove the singularity of the electron's self-energy \cite{Born1934}. In the late 1990s, Deser and Gibbons introduced the Born-Infeld version of gravity theory \cite{Deser1998}, which is a pure metric theory, i.e., the affine connection is given  (a priori) by the Christoffel symbols of the metric. Pure metric Born-Infeld theories lead to fourth order equations and suffer the ghost-like instability in general. Furthermore, the square root determinant form of gravity could trace back to Eddington's  pure affine theory, in which the affine connection is the only dynamical field  on the manifold. Eddington's theory is totally equivalent to general relativity with a cosmological constant \cite{Eddington1924,Schrodinger1950}. Inspired by Eddington gravity, Ba$\tilde{\text{n}}$ados and Ferreira  proposed a new Born-Infeld-like theory called Eddington-inspired Born-Infeld (EiBI) gravity \cite{Banados2010}. Working in the Palatini formalism, in which the metric and connection are regarded as independent fields, the equations of motion are  second order and the ghostlike instabilities can be avoided  \cite{Vollick2004,Delsate2012}. The theory is totally equivalent to general relativity in vacuum but differs from it in the presence of matter. EiBI gravity approaches Eddington's theory in dense or high curvature regions, hence the theory presents some novel properties and modifies the ultraviolet structures of the spacetime.  Especially, the annoying big bang singularities may be avoided in this theory \cite{Banados2010}. Therefore, EiBI gravity has drawn a lot of attention and been widely studied in different topics since its proposal.  The cosmological singularity problems (e.g. big bang singularity, big rip singularity) in this gravity were discussed in Refs. \cite{Pani2011,Escamilla-Rivera2012,Cho2012,Pani2012a,Scargill2012,Yang2013,Jana2013,Kim2014,Bouhmadi-Lopez2014,Bouhmadi-Lopez2015,Bouhmadi-Lopez2014a,Bouhmadi-Lopez2016,Arroja2017,Li2017}. The cosmological and astrophysical constraints were considered in Refs. \cite{Casanellas2012,Avelino2012,Avelino2012b,Sham2014,Banerjee2017}. More cosmological issues, like large scale structure,  inflationary solution and so on, were investigated in Refs. \cite{Du2014,Avelino2012a,Harko2013,Cho2013a,Kim2014a,Harko2014,Cho2014a,Cho2015b,Cho2015a,Cho2015,Avelino2016,Avelino2016a,Afonso2017,Huang2017}. The compact objects were studied in Refs. \cite{Olmo2013,Wei2015,Sotani2014,Jana2015,Sotani2015,Bazeia2017,Harko2013b,Shaikh2015,Olmo2015,Tamang2015,Olmo2016,Sham2012,Sham2013,Harko2013a}.  Some extensions of EiBI theory were presented in Refs.\cite{Delsate2012,Cho2013,Odintsov2014,BeltranJimenez2014,BeltranJimenez2015,Chen2016}.
For an introduction to and summary of Born-Infeld inspired gravities see a recent review \cite{BeltranJimenez2017}  and references therein.

In Refs. \cite{Liu2012,Fu2014}, the authors investigated the thick brane solution in EiBI theory with a scalar field presenting in the five-dimensional background.  The analytic single-kink solution and numerical double-kink solution were achieved.  The transverse-traceless tensor perturbation was studied.  It was shown that the tensor perturbation is stable and the graviton zero mode is localized on the brane, which results in the four-dimensional Newtonian potential. However, it is still not clear whether the scalar perturbations are stable and the scalar zero modes are localized on the brane. It is known that a localized scalar zero mode would lead to an unacceptable four-dimensional long-range force on the brane. Therefore, in order to recover Einstein's general relativity on the brane in the low-energy effective theory, it is required that the scalar perturbations are stable and the scalar zero modes are not localized on the brane.

In this paper, we further investigate the scalar perturbations of the EiBI braneworld solution. In Ref. \cite{Lagos2014}, the authors developed a method to deal with the scalar perturbations of a flat EiBI universe. However, by taking advantage of the bimetric version of EiBI gravity, here we would utilize another convenient way, namely the Arnowitt-Deser-Misner (ADM) decomposition method, to get rid of the redundant degrees of freedom in the scalar perturbations.

The paper is organized as follows. In Sec. \ref{Background},
the background equations of the EiBI branewold model are derived. In Sec. \ref{Perturbations}, the linear scalar perturbations on the EiBI branewold background are considered, and by analyzing the  equations of motion of the physical scalar propagating degree of freedom, the stability condition for scalar perturbations is achieved. In Sec. \ref{Stability}, the stability of an analytic domain wall solution is analyzed.
Finally, conclusions are presented.

 \section{Background Equations}\label{Background}

We start from the bimetric version of the EiBI action with  $g$ the spacetime metric and $q$ the auxiliary metric, which is given by \cite{Delsate2012,Scargill2012,Lagos2014}
\beq
S=\frac{1}{2\kappa}\int d^{d+1}x\left[\sqrt{-q}\left(R(q)-\frac{d-1}{b}\right)
+\frac{1}{b}\left(\sqrt{-q}q^{MN}g_{MN}-2\lambda\sqrt{-g}\right)\right]+S_\text{M}(\varphi,g),
\label{Action_Bimetric}
\eeq
where the Ricci scalar $R(q) \equiv q^{\mu\nu}R_{\mu\nu}$, $\kappa\equiv 8\pi G_5$=1, $d$ refers to the number of spatial dimensions, and $b$ is a constant with inverse dimension to that of the cosmological constant. For the thick brane model, one usually considers  the  background matter to be the standard self-interacting scalar field, i.e.,
\beq
S_\text{M}(\varphi,g)=\frac{1}{2}\int{}d^{d+1}x\sqrt{-g}\lt[-(\nabla\varphi)^2-V(\varphi) \rt].
\eeq

By varying the action with respect to the  metrics $g$ and $q$ respectively, one arrives at the same  equations of motion as in the Palatini formulation \cite{Banados2010}
\beqn
\sqrt{-q}q^{MN}\!&\!=\!&\!\lambda \sqrt{-g} g^{MN}-{ b\sqrt{-g}}T^{MN},
\label{EoM_1}\\
{q_{MN}} \!&\!=\!&\! {g_{MN }} + b{R_{MN }}.
\label{EoM_2}
\eeqn

The background ansatz for the most general metric which preserves $d$-dimensional Poincar\'{e} invariance is
\beq
ds^2=g_{MN}dx^M dx^N=dy^2+a^2(y)\eta_{\mu\nu}dx^\mu dx^\nu,\label{Brane_ST_metric}\\
\eeq
where $a(y)$ is the warp factor.
Thus the corresponding auxiliary metric is given by
\beq
d\tilde{s}^2=q_{MN}dx^M dx^N=X^2(y)dy^2+Y^2(y)a^2(y)\eta_{\mu\nu}dx^\mu dx^\nu.\label{Brane_Aux_metric}
\eeq
To simplify the notation,  we define $Y^2(y)a^2(y)\equiv e^{2\rho(y)}$.
In order to be consistent with the $d$-dimensional
Poincar\'{e} invariance of the metric, we assume that the scalar field depends only on the extra dimension, i.e.,  $\varphi=\varphi(y)$.

With these metrics, Eqs. (\ref{EoM_1}) and (\ref{EoM_2}) give
\beqn
\frac{\lambda }{b} - \frac{{{{\dot \varphi }^2}}}{2} + V-\frac{{{Y^d}}}{{bX}} \!&\!=\!&\! 0,\label{Brane_Eom_1}\\
\frac{\lambda }{b} + \frac{{{{\dot \varphi }^2}}}{2} + V-\frac{{X{Y^{d - 2}}}}{b} \!&\!=\!&\! 0,\label{Brane_Eom_2}\\
1 + (d - 1){X^2} - d\frac{{{X^2}}}{{{Y^2}}} + bd(d - 1){\dot\rho^2} \!&\!=\!& 0,\label{Brane_Eom_3}
\eeqn
where the dot denotes the derivative with respect to the extra dimension $y$.

In order to consider the scalar perturbations around the background brane metrics, it  is more convenient to proceed by working in the ADM formalism \cite{Maldacena2003}.
Because most redundant degrees of freedom act as Lagrange multipliers in this formalism, the physical propagating degrees of freedom are easy to be read off from some nondynamical equations.

The background spacetime metric $g_{MN}$ and auxiliary metric $q_{MN}$ in the ADM formalism are given by
\beqn
ds^2 &=&g_{MN}dx^M dx^N=N^2dy^2+G_{\mu\nu}(dx^\mu+N^\mu dy)(dx^\nu+N^\nu dy),\\
d\tilde{s}^2&=&q_{MN}dx^M dx^N=\,n^2dy^2\,+Q_{\mu\nu}(dx^\mu+n^\mu dy)(dx^\nu+\,n^\nu\,dy).
\eeqn
where
\beqn
N=1,~~N^\mu=0,~~G_{\mu\nu}=a^2(y)\eta_{\mu\nu},~~n=X(y),~~n^\mu=0,~~Q_{\mu\nu}=Y^2(y)a^2(y)\eta_{\mu\nu}.
\eeqn
%The different magnitudes between the two formalisms are
%\beqn
%\sqrt{-g}\!&\!=\!&\!\sqrt{G}N,\\
%\sqrt{-q}\!&\!=\!&\!\sqrt{Q}n.
%\eeqn
%\beq
%(\pt\varphi)^2=g^{MN}\partial_M\varphi\partial_N\varphi=N^{-2}(\dot\varphi-N^\mu\partial_\mu\varphi)^2+G^{\mu\nu}\partial_\mu\varphi\partial_\nu\varphi,
%\eeq
%\beq
%q^{MN}g_{MN}={Q^{\mu\nu}}{G_{\mu\nu}}+\frac{{N^2}}{n^2}-\frac{{G_{\mu\nu}}}{n^2}(2n^\mu N^\nu - N^\mu N^\nu - n^\mu n^\nu).
%\eeq
So in the ADM formalism the bimetric EiBI action (\ref{Action_Bimetric}) is formulated as
\beqn
S &=&\frac{1}{{2 }}\int d^{d+1}{x}\Bigg\{ \frac{{\sqrt {- Q} }}{b}\left[nb{R^{(d)}} -(d-1)n + n{Q^{\mu\nu}}{G_{\mu\nu}} + \frac{{{N^2}}}{n} \right.\nn\\
&&- \left. \left.\frac{b}{n}\left({ E_{\mu\nu}}{ E^{\mu\nu}} - {E^2}\right)- \frac{{{G_{\mu\nu}}}}{n}\left(2{n^\mu}{N^\nu} - {N^\mu}{N^\nu} - {n^\mu}{n^\nu}\right)\right]\right. \nn\\
&&- \sqrt { -G}\left[\frac{{2\lambda }}{b} N + 2NV + {N^{ - 1}}{(\dot \varphi  - {N^\mu}{\partial _\mu}\varphi )^2} + N{G^{\mu\nu}}{\partial _\mu}\varphi {\partial _\nu}\varphi\right] \Bigg\},
\label{Brane_EiBI_ADM_action}
\eeqn
where $R^{(d)}$ and $E_{\mu\nu}$ are
\beqn
R&\!=\!& R^{(d)}-N^{-2}(E_{\mu\nu}E^{\mu\nu}-E^2),\\
E_{\mu\nu}&\!=\!& \frac{1}{2}(\dot Q_{\mu\nu}-\nabla_\mu N_\nu-\nabla_\nu N_\mu),\\
E &\!=\!& E^\mu_\mu.
\eeqn

From the ADM action, it is obvious to see that $Q_{\mu\nu}$ and $\varphi$ are the dynamical variables, while other variables $n$, $N$, $n^\mu$, $N^\mu$ and $G_{\mu\nu}$ are nondynamical and can be regarded as the Lagrange multipliers.  Thus the equations of motion for $n$, $N$, $n^\mu$, $N^\mu$, and $G_{\mu\nu}$ just play the roles of Hamiltonian constraints, which are listed as follows:
\beqn
&&b{R^{(d)}} -(d-1) + \frac{{{b}}}{{{n^2}}}\left({E_{\mu\nu}}{E^{\mu\nu}} - {E^2}\right) + \frac{{{G_{\mu\nu}}}}{{{n^2}}}\left(2{n^\mu}{N^\nu} - {N^\mu}{N^\nu} - {n^\mu}{n^\nu}\right)\nn\\
&&+ {Q^{\mu\nu}}{G_{\mu\nu}} - \frac{{{N^2}}}{{{n^2}}}  = 0.
\label{Brane_Constraint_Eq_1}\\
&&\frac{{\sqrt { -Q} }}{b}\frac{{2N}}{n} - \sqrt { -G} \left[\frac{{2\lambda }}{b} + 2V - \frac{1}{{{N^2}}}{(\dot \varphi  - {N^\mu}{\partial _\mu}\varphi )^2} + {G^{\mu\nu}}{\partial _\mu}\varphi {\partial _\nu}\varphi \right] = 0.
\label{Brane_Constraint_Eq_2}\\
&&{\nabla ^\mu}\left[\frac{b}{n}({E_{\mu\lambda}} - {Q_{\mu\lambda}}E)\right] + \frac{{{G_{\lambda\mu}}}}{n}({N^\mu} - {n^\mu}) = 0.
\label{Brane_Constraint_Eq_3}\\
&&\frac{{\sqrt { -Q} }}{b}\frac{{{G_{\mu\lambda}}}}{n}({n^\mu} - {N^\mu}) - \frac{{\sqrt {-G} }}{N}(\dot \varphi  - {N^\mu}{\partial _\mu}\varphi ){\partial _\lambda}\varphi  = 0.
\label{Brane_Constraint_Eq_4}\\
&&\frac{{\sqrt {  -Q} }}{b}[n{Q^{\mu\nu}} - n^{-1}(2{n^\mu}{N^\nu} - {N^\mu}{N^\nu} - {n^\mu}{n^\nu})]+ \sqrt { - G} N{G^{\mu\lambda}}{G^{\nu\rho}}{\partial _\lambda}\varphi {\partial _\rho}\varphi\nn\\
 &&- \frac{1}{2}\sqrt {  -G} {G^{\mu\nu}}\left[ {\frac{{2\lambda }}{b}N + 2NV + {N^{ - 1}}{{\lt(\dot \varphi  - {N^\lambda}{\partial _\lambda}\varphi \rt)}^2} + N{G^{\lambda\rho}}{\partial _\lambda}\varphi {\partial _\rho}\varphi } \right] = 0.\label{Brane_Constraint_Eq_5}
\eeqn
These equations are obtained from the background metric. By linearly perturbing these background constraints, we will get the constraints for the linear perturbations.

\section{Scalar Perturbations}\label{Perturbations}

It is well known that the linear perturbations around the background metric can be  decomposed into  scalar, transverse vector and transverse-traceless tensor modes (the so-called ``scalar-tensor-vector decomposition")  due to the tensor structure of the equations of motion. After performing the scalar-tensor-vector decomposition, the three kinds of modes decouple with each other. Thus, we will only include the scalar perturbations in the metric. The scalar perturbations in the two metrics are assumed to be
\begin{subequations}\label{Scalar_Perturbations}
\beqn
&&\delta \varphi=\zeta,~\delta N=\Phi,~\delta N_\mu=a^2(y)\pt_\mu\Psi,~\delta G_{\mu\nu}=a^2(y)\lt(2\Xi\eta_{\mu\nu}+\pt_\mu\pt_\nu \Theta\rt),\\
&&\delta n=X(y)\phi,~\delta n_\mu=Y^2(y)a^2(y)\pt_\mu\psi,~\delta Q_{\mu\nu}=Y^2(y)a^2(y)\lt(2\xi\eta_{\mu\nu}+\partial_\mu\partial_\nu \theta\rt).
\eeqn
\end{subequations}

Here we note that the diffeomorphism ensures that the action (\ref{Action_Bimetric})  is invariant under the coordinate transformation $x'^{M}=x^{M}+\epsilon^{M}$ with $\epsilon^M=(\epsilon^\mu,\epsilon^5)$ an arbitrary $(d+1)$-dimensional vector.  In the language of gauge transformations, these perturbations  transform as
\begin{subequations}\label{Gauge_Transformations}
\beqn
\delta \zeta &=&-\dot\varphi \epsilon _5 ,~
\delta\Phi =-\dot\epsilon _5,~
\delta\Psi =a^{-2}\lt(2{\frac{\dot a }{a}\epsilon^s-{\dot \epsilon^s}-\epsilon _5}\rt),~
\delta \Xi =-\frac{\dot a}{a}\epsilon _5 ,~
\delta \Theta =-2a^{-2}\epsilon^s,\\
\delta \phi &=& \dot X\epsilon _5-X \dot \epsilon _5,~
\delta \psi =a^{-2}\lt( 2 \frac{\dot  a}{a} \epsilon^s -{\dot \epsilon^s}-\frac{X^2 }{Y^2}\epsilon _5 \rt),~
\delta \xi=-\dot \rho \epsilon _5,~
\delta \theta =-2a^{-2} \epsilon^s,
\eeqn
\end{subequations}
where $\epsilon_\mu=\pt_\mu\epsilon^s+\epsilon^V_\mu$ with $\pt^\mu\epsilon^V_\mu=0$,  and $\epsilon_M=g_{MN}\epsilon^N$. Here $\epsilon ^s$ and $\epsilon_5$ are two arbitrary infinitesimal functions, thus there are two gauge freedoms in these scalar perturbations. Now in order to fix the gauge freedoms, we work in the unitary gauge, i.e., we choose $\epsilon_5$ and $\epsilon^s$ to set $\zeta=\theta=0$. This is because these two perturbations are related to dynamical variables $Q_{MN}$ and $\varphi$. After gauge fixing, the only dynamical perturbation is $\xi$, whose equation of motion can be obtained from the perturbed equations.

Then, the linear perturbation of Eq. (\ref{Brane_Constraint_Eq_1}) gives
\beqn
- \frac{{b(d - 1)}}{{{a^2}}}{\partial ^2}\xi  + d(\Xi  - \xi )  - \frac{{{Y^2}}}{{{X^2}}}\Phi  + \lt(d - (d - 1){Y^2}\rt)\phi\nn\\
+ b(d - 1)\frac{{{Y^2}}}{{{X^2}}}\dot\rho({\partial ^2}\psi  - d{\dot \xi })+\frac{1}{2}{\partial _\mu}{\partial ^\mu}\Theta=0.
\label{Brane_Pert_Constraint_Eq_1}
\eeqn
The linear perturbations of Eqs.  (\ref{Brane_Constraint_Eq_2}), (\ref{Brane_Constraint_Eq_3}) and (\ref{Brane_Constraint_Eq_4}) give
\beqn
\phi-d(\xi-\Xi) - \left(2-\frac{X^2}{Y^2}\right)\Phi+\frac{1}{2}{\partial _\mu}{\partial ^\mu}\Theta&=&0,
\label{Brane_Pert_Constraint_Eq_2}\\
{\partial _\lambda}\left[\dot\rho\phi  -\dot \xi  + \frac{{{a^2}}}{b(d - 1)}\left(\Psi  - \psi \right)\right] &=& 0,
\label{Brane_Pert_Constraint_Eq_3}\\
{\partial _\lambda}(\psi  - \Psi ) &=&0.
\label{Brane_Pert_Constraint_Eq_4}
\eeqn
The perturbed part proportional to $\delta^{\mu\nu}$  of the constraint  equation(\ref{Brane_Constraint_Eq_5}) gives
\beqn
\phi+(d-2)(\xi-\Xi)-\frac{Y^2}{X^2}\Phi- \frac{1}{2}{\partial _\mu}{\partial ^\mu}\Theta=0.
\label{Brane_Pert_Constraint_Eq_5}
\eeqn
The other part of the constraint equation (\ref{Brane_Constraint_Eq_5}) in the form of $\pt^\mu\pt^\nu S$ (where $S$ is any scalar) simply gives
\beq
{\partial ^\mu}{\partial ^\nu}\Theta=0.
\label{Brane_Pert_Constraint_Eq_6}
\eeq

The above equations give the Hamiltonian constraints for the scalar perturbations, and these constraints ensure that one can eliminate the remaining nonphysical degrees of freedom.  On the other hand, since the matter is covariantly coupled to the metric $g$, the matter conservation equation $\nabla_{M} T^{MN}=0$ holds, where the covariant derivative refers to the spacetime metric $g$. The conservation equation leads to a scalar field equation $\Box^{(d+1)}\varphi-dV(\varphi)/d\varphi=0$. So up to first order perturbation, the scalar field equation gives
\beq
{\partial ^2}\Psi  - d {\dot\Xi} + {\dot\Phi } + 2\lt(d\frac{{{\dot a }}}{a} + \frac{\ddot\varphi }{\dot\varphi}\rt)\Phi =0.
\label{Brane_Perturbed_KG_Eq}
\eeq

First,  from the constraint equations (\ref{Brane_Pert_Constraint_Eq_4}) and (\ref{Brane_Pert_Constraint_Eq_6}), we have
\beq
\Theta=0, ~~ \Psi=\psi.
\eeq
Then Eq. (\ref{Brane_Pert_Constraint_Eq_3}) simply gives
\beq
\phi  = \frac{\dot \xi}{\dot\rho}.
\label{Brane_relation_phi_xi}
\eeq
Further, from Eqs. (\ref{Brane_Pert_Constraint_Eq_1}) and (\ref{Brane_Pert_Constraint_Eq_5}), we have
\beq
{\partial ^2}\Psi  = d{\dot\xi } + \frac{{{X^2}}}{{{a^2}{Y^2}{\dot\rho }}}{\partial ^2}\xi  + \frac{{{X^2}}}{{b{Y^2}{\dot\rho }}}\lt[\lt({Y^2} - 1\rt)\phi + 2(\xi  - \Xi )\rt].
\label{partial2Psi}
\eeq
By substituting Eq. (\ref{partial2Psi}) into Eq. (\ref{Brane_Perturbed_KG_Eq}), we get
\beq
\frac{{{X^2}}}{{{a^2}{Y^2}{\dot \rho }}}{\partial ^2}\xi  + d\lt({\dot \xi  } - {\dot \Xi  }\rt) + 2\frac{{{X^2}}}{b{Y^2}{\dot \rho } }(\xi  - \Xi ) + \frac{{{X^2}}}{{b{Y^2}{\dot \rho  }}}\lt({Y^2} - 1\rt)\phi + {\dot \Phi } + 2\Phi \Big(d\frac{{{\dot a }}}{a} + \frac{{\ddot \varphi}}{{\dot \varphi }}\Big)=0.
\eeq
Then eliminating $\xi  - \Xi$ with Eq. (\ref{Brane_Pert_Constraint_Eq_5}), we have
\beqn
&&\frac{{{X^2}}}{{{a^2}{Y^2}{\dot \rho}}}{\partial ^2}\xi  - \frac{d}{{d - 2}}{\dot \phi} + \frac{{{X^2}}}{{b{Y^2}{\dot \rho  }}}\left({Y^2} - \frac{d}{{d - 2}}\right)\phi + \left(1 + \frac{d}{{d - 2}}\frac{{{Y^2}}}{{{X^2}}}\right){\dot \Phi  } \nn\\
&& + 2\lt(d\frac{{{\dot a }}}{a} + \frac{{\ddot \varphi}}{{\dot \varphi }} + \frac{1}{{d - 2}}\frac{1}{{b{\dot \rho }}} + \frac{d}{{d - 2}}\frac{{{Y^2}}}{{{X^2}}}\lt(\frac{{{\dot Y}}}{Y} - \frac{{{\dot X}}}{X}\rt)\rt)\Phi=0.
\eeqn
Moreover, the combination of Eqs. (\ref{Brane_Pert_Constraint_Eq_2}) and (\ref{Brane_Pert_Constraint_Eq_5}) gives
\beq
\Phi={2 (d - 1)\phi}/{F_0},
\eeq
where $F_0\equiv2(d - 2) - (d - 2)\frac{{{X^2}}}{{{Y^2}}} + d\frac{{{Y^2}}}{{{X^2}}}$.

So after eliminating $\Phi$, the perturbed equation can be rewritten as
\beqn
&&\lt[\frac{X^2}{{b{Y^2}{\dot \rho }}}\lt({Y^2} - \frac{d}{{d - 2}}\rt) - 2(d - 1)\lt(1 + \frac{d}{{d - 2}}\frac{Y^2}{X^2}\rt)\frac{{\dot F_0}}{F_0^2}\rt. \nn\\
&&\lt.+ \frac{{4(d - 1)}}{{F_0}}\lt(d\frac{{{\dot a }}}{a} + \frac{\ddot\varphi }{\dot\varphi} + \frac{1}{d - 2}\frac{1}{b{\dot \rho}} + \frac{d}{{d - 2}}\frac{Y^2}{X^2}\lt(\frac{\dot Y }{Y} - \frac{{{\dot X}}}{X}\rt)\rt)\rt]\phi \nn\\
&&- \lt[\frac{d}{{d - 2}} - \frac{{2(d - 1)}}{F_0}\lt(1 + \frac{d}{{d - 2}}\frac{Y^2}{X^2}\rt)\rt]{\dot \phi}+\frac{{{X^2}}}{{{a^2}{Y^2}{\dot \rho  }}}{\partial ^2}\xi =0.
\eeqn

Finally, by utilizing the relation (\ref{Brane_relation_phi_xi}) to eliminate $\phi$, we arrive at the expected dynamical equation with only one physical propagating degree of freedom $\xi$, i.e.,
\beq
F_1(y) {\ddot \xi} + F_2(y){\dot \xi  } + {\partial ^2}\xi=0,
\label{Scalar_Pert_EQ}
\eeq
where
\beqn
F_1(y)\!&\!=\!&\!- \frac{a^2 Y^2}{X^2}\lt[\frac{d}{{d - 2}} - \frac{2(d - 1)}{F_0}\lt(1 + \frac{d}{{d - 2}}\frac{{{Y^2}}}{{{X^2}}}\rt)\rt],\label{Exp_F1}\\
F_2(y)\!&\!=\!&\!- F_1\frac{\ddot \rho }{\dot \rho}+\frac{a^2}{b{\dot \rho }}\lt({Y^2} - \frac{d}{d - 2}\rt)- \frac{{{a^2}{Y^2}}}{{{X^2}}}\lt[2(d - 1)\lt(1 + \frac{d}{{d - 2}}\frac{{{Y^2}}}{{{X^2}}}\rt)\frac{\dot F_0}{F_0^2}\rt.\nn\\
 &&\lt.- \frac{{4(d - 1)}}{F_0}\lt(d\frac{{{\dot a }}}{a} + \frac{\ddot \varphi}{\dot \varphi } + \frac{1}{{d - 2}}\frac{1}{b{\dot \rho  }} + \frac{d}{{d - 2}}\frac{{{Y^2}}}{{{X^2}}}\lt(\frac{\dot Y  }{Y} - \frac{\dot X}{X}\rt)\rt)\rt].
\eeqn

We recall that diffeomorphism generates the gauge transformation invariance as shown in (\ref{Gauge_Transformations}), where  $\xi$ and $\zeta$ transform as $\delta\xi=-\dot\rho\epsilon_{5}$ and $\delta\zeta=-\dot\varphi\epsilon_{5}$. Thus, one can construct a gauge-invariant combination $\mathcal{R}=\xi-\fc{\dot\rho}{\dot\varphi}\zeta$. Since we work in unitary gauge, where $\varphi$ is frozen to its background value $\delta\varphi=\zeta=0$, the gauge-invariant variable $\mathcal{R}$ is just identical to the metric perturbation $\xi$. Therefore, Eq. (\ref{Scalar_Pert_EQ}) describes the scalar perturbation in a gauge-independent way.

In order to analyze the stability under the scalar perturbation, we decompose $\xi(x,z)$ as
\beq
\xi(x,y)=\tilde\xi(x)\chi(y).
\eeq
Because of the manifest $d$-dimensional Poincar\'{e}  invariance in the metric (\ref{Brane_ST_metric}), the field $\tilde\xi(x)$ satisfies the $d$-dimensional Klein-Gordon equation ${\partial ^2}\tilde\xi(x)=m_n^2\tilde\xi(x),$ with  $m_n$ being the observed $d$-dimensional effective mass of the scalar KK excitations $\tilde\xi(x)$. This is the so-called KK decomposition.
Then Eq. (\ref{Scalar_Pert_EQ}) is rewritten as
\beq
F_1 (y){\ddot \chi} + F_2(y){\dot \chi  } +m_n^2\chi=0.
\label{Scalar_Pert_EQ_2}
\eeq

In order to eliminate the prefactor of the second derivative term $F_1$, for $F_1(y)>0$ we make a coordinate transformation as $dz=dy/\sqrt{F_1}$, then Eq. (\ref{Scalar_Pert_EQ_2}) is rewritten as
\beq
\frac{{{d^2}\chi(z) }}{{d{z^2}}} +F_3 (z)\frac{{d\chi(z) }}{{dz}} + m_n^2\chi(z)  = 0, \label{Eq_withF1F3}
\eeq
where $F_3(z)\equiv\frac{F_2}{\sqrt {F_1} }-\frac{F_1'}{2F_1}$ with the prime denoting the derivative with respect to the extra dimension coordinate $z$.

Further, to eliminate the first derivative term in Eq. (\ref{Eq_withF1F3}), we decompose $\chi$ as
\beq
\chi(z)=e^{-\int\fc{F_3}{2}dz}\Upsilon(z).
\eeq
Then we arrive at a Schr\"{o}dinger-like equation
\beq
-\Upsilon''(z)+  V(z) \Upsilon(z)=m_n^2\Upsilon(z),
\eeq
where the effective potential $V(z)$ is given by
\beq
V(z)=\frac{{F_3'}}{2} + \frac{{{F_3^2}}}{4}.
\label{Potential}
\eeq

This Hamiltonian can be factorized as a supersymmetric quantum mechanics form
\beq
H= -\pt_z^2+V(z)=A^\dag A=\lt(\pt_z+\fc{F_3}{2}\rt)\lt(-\pt_z+\fc{F_3}{2}\rt).
\eeq
It is easy to see that the eigenvalues of $H$ are non-negative for the Newmann boundary condition, i.e.,
\beqn
m_n^2\int dz|\Upsilon(z)|^2&=&\int dz \Upsilon^\dag H\Upsilon=\int dz\Upsilon^\dag \lt(\overrightarrow{\pt}_z+\fc{F_3}{2}\rt)\lt(-\overrightarrow{\pt}_z+\fc{F_3}{2}\rt)\Upsilon
\nn\\
&=&\int dz\Upsilon^\dag \lt(-\overleftarrow{\pt}_z+\fc{F_3}{2}\rt)\lt(-\overrightarrow{\pt_z}+\fc{F_3}{2}\rt)\Upsilon+\int dz \pt_z\lt[\Upsilon^\dag \lt(-\overrightarrow{\pt_z}+\fc{F_3}{2}\rt)\Upsilon\rt]\nn\\
&=&\int dz|A\Upsilon|^2+\lt.\Upsilon^\dag A\Upsilon\rt|_{\text{Boundary}}.
\eeqn
For the Neumann boundary condition $\lt.\pt_z\xi(x,z)\rt|_{\text{Boundary}}=0$, which simply reduces to $\lt.A\Upsilon\rt|_{\text{Boundary}}=0$, the eigenvalues are non-negative, $m_n^2\geq0$. So the system is stable under scalar perturbations.

However, for $F_1(y)<0$ the coordinate transformation is $dz=dy/\sqrt{-F_1}$, so the Schr\"{o}dinger-like equation is given by
\beq
\tilde H\Upsilon(z)=\lt(\pt_z+\fc{\tilde F_3}{2}\rt)\lt(-\pt_z+\fc{\tilde F_3}{2}\rt)\Upsilon(z)=-m_n^2\Upsilon(z),
\eeq
where $\tilde F_3(z)=\frac{-F_2}{\sqrt {-F_1} }-\frac{F_1'}{2F_1}$.
Then the self-adjoint Hamiltonian gives non-negative eigenvalues $-m_n^2\geq0$, i.e., $m_n^2\leq0$. Thus there are tachyonic modes, and the system is unstable under the scalar perturbations.

If $F_1$ vanishes, Eq. (\ref{Exp_F1}) gives $X^2={(1\pm\sqrt{1-d^2})Y^2}/{d}$. This implies that there is no real root. However, the spacetime metric must be real, so this case is excluded.

In summary, the sufficient condition for the system to be stable under linear scalar perturbations is $F_1(y)>0$.

\section{Stability of EiBI brane solutions}\label{Stability}

An analytic  domain wall brane solution of five-dimensional EiBI gravity was given by Refs. \cite{Liu2012,Fu2014}, where the solution is read as
\beqn
a(y)&=& \text{sech}^{\frac{3}{{4p}}}\lt(ky\rt), ~~(p>0), \\
\varphi'(y)&=& K a^{2p}(y),  \\
{X^2(y)} &=& \lt (bK^2\rt)^{\fc{2}{3}} \lt(\frac{{p + 1}}{{\sqrt p }}\rt)^{\fc{4}{3}}{a^{\frac{{8p}}{3}}(y)} ,\\
{Y^2(y)}  &=& \lt(bK^2\rt)^{\fc{2}{3}} \lt(\frac{{p + 1}}{{{p^2}}}\rt)^{\fc{1}{3}}{a^{\frac{{8p}}{3}}(y)},
\eeqn
where the parameters  $b>0$, $k=\frac{{2p}}{{\sqrt {3b(3 + 4p)} }}$ and  $K = \pm\frac{{{{(1 + 4p/3)}^{3/4}}}}{{p + 1}}\sqrt {\frac{p}{b}}$. It has been shown that the tensor perturbation is stable and the graviton zero mode can be localized on the brane for any positive $p$ \cite{Liu2012,Fu2014}, which will result in the four-dimensional Newtonian potential.

With this solution, $F_1(y)$ is calculated as
\beq
F_1(y)=\frac{\lt(8 + 8p + 5{p^2}\rt)}{(1 + p)\lt(8 - 3{p^2}\rt)}a^2(y).
\eeq
It is clear that $F_1(y)$ is positive if and only if $0<p<\sqrt{8/3}$. Thus, only the solution with $0<p<\sqrt{8/3}$ is tachyonic-free and stable under scalar perturbations.

%%%%%%%%%%%%%%%%%%%%%%%%%%%%%%%%%%%%%%%%%%%%%
\begin{figure*}[htb]
\begin{center}
\includegraphics[width=7cm,height=5cm]{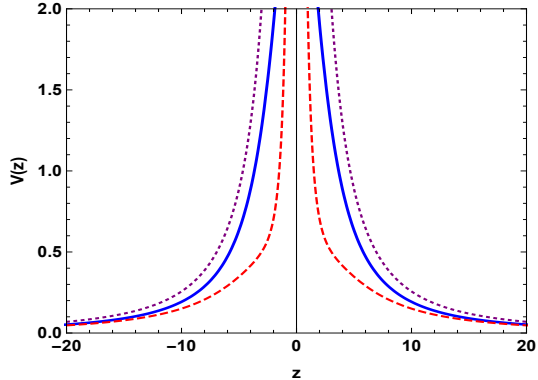}
\end{center}
\caption{The potential $V(z)$ with the
parameters $p=\sqrt{2}, b=1$ (dotted line), $p=1, b=1$ (thick line),  and $p=1, b=5$ (dashed line).}
\label{Fig_Potential}
\end{figure*}
%%%%%%%%%%%%%%%%%%%%%%%%%%%%%%%%%%%%%%%%%%%%%

Furthermore, in order to recover the familiar four-dimensional gravity at low energy, the scalar zero mode (i.e., $m=0$) should not be localized on the brane, otherwise it would lead to an unacceptable long-range force. 
We show the potential, Eq. (\ref{Potential}), of the Schr\"{o}dinger-like equation in Fig. \ref{Fig_Potential} for some values of  parameters $p$ and $b$ as examples.  The potential is convex and positive everywhere,  and approaches zero  when $|z|\rightarrow\infty$. Thus, the spectrum is continuous and starts from $m^2>0$. Especially, the potential blows up at the origin, because $\dot\rho(0)=0$  which appears in the denominator of $F_3$. Because of the infinite barrier, all the eigenfunctions will be suppressed to zero at the origin and turn into plane waves where they are far away from the brane. So although the potential is singular, the wave function is regular everywhere.  Any scalar perturbations will be totally reflected back to infinity.  Therefore, none of the scalar modes are localized on the brane and they will not contribute to the interaction of the particles on the brane at low energy.

\section{Conclusions}

In this paper, we have investigated the linear scalar perturbations of the EiBI braneworld model using the ADM decomposition method, which is proved to be a convenient way to fix the gauge freedoms and to remove the nonphysical degrees of freedom in this theory.  The application in cosmological perturbations is just straightforward. After some cumbersome but simple algebra, the equation of  motion for the physical perturbation $\xi$ was achieved. Further, with the  KK decomposition, we obtained a Schr\"{o}dinger-like equation with mass square of the KK excitations as the eigenvalue.  It was shown that the stability condition of the linear scalar perturbations for the EiBI braneworld model is $F_1(y)>0$. Finally, the stability of an analytic domain wall solution was analyzed under this criterion. We found that only the solution with $0<p<\sqrt{8/3}$ is stable under linear scalar perturbations and there is no unacceptable new long-range force in this model.

We have shown that the ADM decomposition method is useful for dealing with the scalar perturbation of EiBI theory. Actually, this method is also applicable for more general Palatini theories. Here we take the Palatini $f(\mathcal{R})$ theory as an example, where $\mathcal{R}\equiv g^{MN}\mathcal{R}_{MN}(\Gamma)$. By introducing an auxiliary metric $q_{MN}=\phi^{2/3}g_{MN}$ with $\phi\equiv{df(\mathcal{R})}/{d\mathcal{R}}$, which is compatible to the affine connection $\Gamma$, the theory can be expressed as a bimetric version. Further, by rewriting $\mathcal{R}\lt(q(g,\phi)\rt)=R(g)+\cdots $ and imposing a conformal transformation $g_{MN}\rightarrow \tilde{g}_{MN}=\phi^{2/3}g_{MN}$, one arrives at the well-known Einstein frame of $f(\mathcal{R})$ theory, in which a Ricci scalar minimally couples to a ``new" scalar degree of freedom $\phi$ \cite{Sotiriou2010}. Now it is straightforward to apply the ADM decomposition method. The scalar perturbation for braneworld models in more general Palatini theories, such as $f\lt(\mathcal{R}_{(MN)}(\Gamma), g^{MN}\rt)$ \cite{BeltranJimenez2015}, is left for our future work.

\section*{ACKNOWLEDGMENTS}

We thank Qi-Ming Fu and Yi Zhong for helpful discussions. This work was supported by the National Natural Science Foundation of China under Grant No. 11522541 and No. 11375075.
K. Yang acknowledges the support of ``Fundamental Research Funds for the Central Universities" under Grant No. SWU-116052. Y.X. Liu acknowledges the support of ``Fundamental Research Funds for the Central Universities" under Grant No. lzujbky-2016-k04.

%\section*{References}

\providecommand{\href}[2]{#2}\begingroup\raggedright\endgroup

\end{document}